\documentstyle[epsfig,aps,floats,preprint,floats]{revtex}

\newcommand{\pa}{\partial} \newcommand{\co}{\nabla}
\newcommand{\vphi}{\varphi} 

\def\bard#1{\setbox0=\hbox{\it /}\wd0=0pt\box0#1}

\newcommand{\beq}{\begin{equation}} 
\newcommand{\eeq}{\end{equation}}
\newcommand{\bea}{\begin{eqnarray}} 
\newcommand{\eea}{\end{eqnarray}}
\newcommand{\beam}{\begin{mathletters}} 
\newcommand{\eeam}{\end{mathletters}}

\begin{document}
\draft
\title{\Large\bf Effective action and tension renormalization 
for cosmic and fundamental strings} 
\author{Alessandra Buonanno$^1$ and
Thibault Damour$^{1,2}$}
\address{$^1$ {\it Institut des Hautes Etudes Scientifiques, 91440
Bures-sur-Yvette, France} \\ 
{$^2$ {\it DARC, CNRS-Observatoire de Paris, 92195 Meudon, France}}}
\vskip 1.5truecm 
\maketitle
\begin{abstract}
We derive the effective action for classical strings 
coupled to dilatonic, gravitational, and axionic fields. 
We show how to use this effective action for: (i) renormalizing the string 
tension, (ii) linking ultraviolet divergences to the infrared (long-range)
interaction between strings, (iii) bringing additional light 
on the special cancellations that occur for fundamental strings, 
and (iv) pointing out the limitations of Dirac's celebrated 
field-energy approach to renormalization.
\end{abstract}

In many elementary particle models cosmic strings are expected 
to form abundantly at phase transitions in the early universe 
\cite{VilenkinShellard94}, \cite{HK}. Oscillating loops of cosmic string might 
be a copious source of the various fields or quanta to which 
they are coupled. They might generate observationally significant 
stochastic backgrounds of: gravitational waves \cite{Vilenkin1981d}, 
massless Goldstone bosons \cite{Davis1985a}, light axions 
\cite{Davis1986}, \cite{DavisShellard1989c}, or light dilatons 
\cite{DamourVilenkin97} (for recent references on stochastic 
backgrounds generated by cosmic strings, see the reviews 
\cite{VilenkinShellard94}, \cite{HK}).  
An oscillating loop which emits outgoing gravitational, axionic or dilatonic waves, 
will also self-interact with the corresponding fields it has generated. 
This self-interaction is formally infinite if the string is modelled 
as being infinitely thin. Such 
infinite self-field situations are well known in the context of 
self-interacting particles. 
It was emphasized long ago by Dirac\cite{Dirac38}, in the case of a 
classical point-like electron moving in its own electromagnetic field, 
that the infinite self interaction problem is cured by renormalizing the mass: 
\beq
\label{1}
m(\delta) = m_R - \frac{e^2}{2\delta}\,,
\eeq
where $m(\delta)$ is the (ultraviolet divergent) bare mass of the electron,
$m_R$ the renormalized mass and $\delta$ a cutoff radius around the electron. 
The analogous problem for self-interacting cosmic strings
has been studied in Refs~\cite{LundRegge}, \cite{DQ90}, \cite{BS}
for the coupling to the axion field, in Ref.~\cite{QuashnockSpergel90} 
for the coupling to the gravitational field, and in Ref.~\cite{CHH90} for the coupling 
to the gravitational, dilatonic and axionic fields.
See also Ref.~\cite{C97} for the coupling to the electromagnetic 
field, in the case of superconducting strings. 
Related work by Dabholkar et al. \cite{DH89}, \cite{DGHR90} pointed 
out the remarkable cancellations, between the dilatonic, gravitational, 
and axionic self-field effects, which take place for (macroscopic)
 fundamental strings. Though these cancellations can be derived for 
superstrings by appealing to supersymmetry (and the existence of 
string-like BPS states \cite{DGHR90}), they also take place 
for bosonic strings. It seems therefore useful to 
deepen their understanding without appealing to supersymmetry. 

The analog of the linearly-divergent renormalization (\ref{1})
of the mass of a point particle is, for a string (in four-dimensional 
spacetime), a logarithmically-divergent 
renormalization of the string tension $\mu$, of the general form 
\beq
\label{3}
\mu (\delta) = \mu_R + C\,\log \, \left (\frac{\Delta_R}{\delta} \right )\,, 
\quad \quad C=C_\vphi + C_g+ C_B\,.
\eeq
The renormalization coefficient $C$ is a sum of contributions 
due to each (irreducible) field with which the string interacts. 
As above $\delta$ denotes the ultraviolet cutoff length, while $\Delta_R$ 
denotes an arbitrary renormalization length which must be introduced 
because of the logarithmic nature of the ultraviolet divergence. 

In this paper, we revisit the problem of the determination of the renormalization 
coefficient $C$ (which, as we shall see, has been heretofore uncorrectly
treated in the literature) with special emphasis on:
(i) the streamlined extraction of $C$ from the one-loop 
(quantum and classical) effective action for self-interacting 
strings, namely ($\alpha$ and $\lambda$ denoting, respectively, the scalar 
and axionic coupling parameters; see Eq.(\ref{4}) below)
\beam
\bea
\label{17a}
&& C_\vphi^{\rm effective-action} = +4 \, \alpha^2 \, G \, \mu^2 \,,\\
\label{17b}
&& C_g^{\rm effective-action} = 0 \,,\\
\label{17c}
&& C_B^{\rm effective-action} =-4\,G \, \lambda^2\,,
\eea
\eeam
(ii) the link between the ultraviolet 
divergence (\ref{3}) and the infrared (long-range) interaction between 
strings, (iii) the special cancellations that occur in $C$ 
for fundamental (super)-strings \cite{DH89}, \cite{DGHR90}, 
and (iv) the fact that the seemingly ``clear'' connection, 
pointed out by Dirac, between renormalization and field energy is valid 
only for electromagnetic and axionic fields but fails to give the correct sign 
and magnitude of $C$  for gravitational and scalar fields. 

In an independent paper, based on a quite different tensorial formalism, 
\cite{C89+93}, \cite{BattyeCarter}, Carter and Battye \cite{CB}, have reached  
conclusions consistent with ours for what concerns the vanishing of the
gravitational contribution $C_g$. [We shall not consider here the finite 
``reactive'' contributions to the equations
of motion which remain after renormalization of the tension 
(see \cite{DQ90}, \cite{BS}, \cite{BD98}).] 

The present work has been motivated by several puzzles 
concerning the various contributions to the renormalization
coefficient $C$. First, Ref.~\cite{DH89}
worked out the three contributions to the classical field 
energy around a straight (infinite) fundamental string and found a cancellation 
between two {\it positive and equal } contributions 
due to $\vphi$ and $B$ and a {\it doubled negative} contribution from 
gravity. We recall that Dirac emphasized that the cutoff dependence 
of the bare electron mass $m(\delta)$ (for a fixed observable mass $m_R$) 
was compatible with the idea that $m(\delta)$ represents the total 
mass-energy of the particle plus that of the electromagnetic field 
contained within the radius $\delta$, so that: 
\beq
\label{2}
m(\delta_2) - m(\delta_1) = + 
\int_{\delta_1}^{\delta_2}d^3x\,T^{00}_{\rm field}\,,
\eeq
with $T^{00}_{\rm field} = E^2/(8\pi)=e^2/(8\pi\,r^4)$.
If we were to apply Dirac's seemingly general result (\ref{2}), 
the work of Ref.~\cite{DH89} (generalized to arbitrary couplings $\alpha$, 
$\lambda$) would be translated into the following ``field-energy'' values 
of the renormalization coefficients:
\beam
\bea
\label{18a}
&& C_{\vphi\,\, \rm expected}^{\rm field-energy} = -4\alpha^2\,G\,\mu^2\,, \\
\label{18b}
&& C_{g\,\,\rm expected}^{\rm field-energy} = +8G\,\mu^2\,,\\
\label{18c}
&& C_{B\,\,\rm expected}^{\rm field-energy}=-4G\,\lambda^2\,.
\eea
\eeam
Only $C_B^{\rm field-energy}$ agrees with $C_B^{\rm effective-action}$ above.  
The sign of $C_\vphi^{\rm field-energy}$ is wrong, as well as the value 
of $C_g^{\rm field-energy}$. Yet, the three partial $C$'s correctly 
cancel in the case of fundamental strings! (See Eq.~(\ref{6}) below).
A second (related) aspect of Eqs.~(\ref{17a})--(\ref{17c}) which needs to be understood 
concerns the vanishing of the gravitational contribution $C_g^{\rm effective-action}$.
Is this an accident or is there a simple understanding of it? 
A further puzzle is raised by the fact that the (nonvanishing) value
(\ref{18b}) for $C_g$ was reproduced by the dynamical calculation of Ref.~\cite{CHH90}.

To answer these puzzles we have computed the effective action 
obtained by eliminating to first order (in a weak field expansion) 
the fields in the total 
action. To clarify the physical meaning of this effective 
action (at both the quantum and classical levels) let us consider 
a generic action of the form 
\beq
\label{19}
S_{\rm tot}[z,A] = 
S_0^{\rm system}[z] - \frac{1}{2} A P^{-1} A + J A\,, 
\eeq
where $P^{-1}$ is the inverse of the propagator of the field 
$A$ (after suitable gauge fixing), and where $J[z]$ is the source of $A$ 
(which depends on the dynamical system described by the variables $z$). 
We use here a compact notation which suppresses both integration over 
spacetime and any (Lorentz or internal) labels on the fields:
e.g. $J A \equiv \int d^nx\,J^i(x) A_i(x)$. 
The quantum effective action for the dynamical system $z$ arises 
when one considers processes where no real field quanta are emitted 
\cite{Feynman50}. It is defined by integrating out the $A$ field with 
trivial boundary conditions at infinity, namely
\beq
\label{21}
\exp i S_q^{\rm eff}[z] = \langle 0_A^{\rm out}|0_A^{\rm in}\rangle_z = 
\int {\cal D}A\,\exp \left (i\left [ S_0 - \frac{1}{2}A P^{-1} A +
J A\right ]\right) = 
\exp i\left [ S_0 + \frac{1}{2}J P_F J\right]\,,
\eeq
where the integration (being Gaussian) is equivalent to estimating 
the integrand at the saddle-point, $\delta S_{\rm tot}/\delta A_0
=-P^{-1} A_0 + J=0$, and where, as is well known \cite{Feynman50}, 
\cite{Weinberg}, the trivial euclidean 
boundary conditions (or the vacuum-to-vacuum prescription) translate 
into the appearance of the Feynman propagator.  For massless fields 
in Feynman-like gauges,  
we can write
\beq
\label{22} 
P_F(x,y) = \int \bard{d} p \,e^{ip(x-y)}\,\frac{R}{p^2-i\epsilon}\,,
\eeq
where $\bard{d} p ={d^np}/{(2\pi)^n}$ and 
$R$ (the {\it residue} of the propagator) is a momentum-independent 
matrix $R_{ij}$, when the field comes 
equipped with a (Lorentz or internal) label: $A_i$. The real part 
of the quantum effective action, $\mbox{Re}[S_q^{\rm eff}[z]]\equiv S^{\rm eff}_c[z]$, 
reads
\beq
\label{20}
S^{\rm eff}_c[z] = S_0^{\rm system}[z] +S_1[z]\,,\,\,
S_1[z] = \frac{1}{2}J[z] P_{\rm sym} J[z]\,, 
\eeq
\beq
\label{Re}
P_{\rm sym}\equiv \mbox{Re}[P_F]=
 \int \bard{d} p\,e^{ip(x-y)}\,\mbox{PP}\left (\frac{R}{p^2}\right )\,,
\eeq
with $\mbox{PP}$ denoting the principal part. 
$S_c^{\rm eff}$ corresponds to a phase difference between 
the in-$A$-vacuum $|0^{\rm in}_A\rangle$ and the out-$A$-vacuum $|0^{\rm out}_A \rangle$.
On the other hand, twice the imaginary part of $S_q^{\rm eff}[z]$ gives the 
probability for the vacuum to remain vacuum: 
$|\langle 0_A^{\rm out}|0_A^{\rm in}\rangle|^2=\exp(-2 \mbox{Im}S_q^{\rm eff})$, 
and is equal to the mean number of $A$-quanta emitted, 
\beq
\bar{n}_A= 2\mbox{Im} S_q^{\rm eff}= \pi \int \bard{d} p 
\delta(p^2) J(-p) R J(p)\,,
\eeq 
where $J(p)\equiv \int d^n x  e^{-ipx} J(x)$ \cite{IZ}.

It is easily checked that $P_{\rm sym}$ is nothing but the classical 
{\it symmetric}, half-retarded--half-advanced propagator. 
This shows that $\mbox{Re}[S_q^{\rm eff}]$ is the {\it classical} 
effective action, obtained by eliminating the field $A$ in (\ref{19}) 
by using the field equations written in the context of a classical 
non-dissipative system, i.e. a system interacting via 
half-retarded--half-advanced potentials. [In the case of 
interacting point charges $S_c^{\rm eff}$ is the Fokker-(Wheeler-Feynman) action.]
Written more explicitly, the ``one-classical-loop'' (i.e. one classical self-interaction)
contribution $S_1$ in Eq.~(\ref{20}) reads 
\bea
S_1[z] &=& \frac{1}{2}\int \int d^nx \,d^ny\,J^i(x) P^{\rm sym}_{ij}(x,y) J^j(y)\nonumber \\
\label{23}
&=& \frac{1}{2}\int \int d^nx \,d^ny\,G_{\rm sym}(x,y)\,J^i(x) R_{i j} J^j(y)\,,
\eea
where we used, from Eq.~(\ref{Re}), $P_{ij}^{\rm sym}(x,y)=R_{ij} G_{\rm sym}(x,y)$, 
$G_{\rm sym}$ being the symmetric scalar Green function: $\Box G_{\rm sym}(x,y)= 
-\delta^{(n)}(x-y)$.
It is easily checked ({\it a posteriori}) that varying with respect to the system 
variables $z$ the classical effective action $S_0[z]+S_1[z]$ 
reproduces the correct equations of motion 
$\delta S_{\rm tot}[z,A_{\rm sym}]/\delta z=0$ with $A_{\rm sym}= P_{\rm sym} J$
being the classical half-retarded--half-advanced potential. 

Let us now apply this general formalism to string dynamics. 
We consider a closed Nambu-Goto string $z^\mu(\sigma^a)$ 
(with $\sigma^a=(\sigma^0, \sigma^1)$) interacting with 
gravitational $g_{\mu \nu}(x^\lambda) = \eta_{\mu \nu} 
+ h_{\mu \nu}(x^\lambda)$, dilatonic $\vphi(x)$ and axionic (Kalb-Ramond)
$B_{\mu \nu}(x)$ fields.  The action for this system is 
$S_{\rm tot} = S_s + S_f$, where a generic action for the string 
coupled to $g_{\mu \nu}$, $\vphi$ and $B_{\mu \nu}$ reads 
\beq
\label{4} 
S_s = - \mu\,\int e^{2\alpha \vphi}\,
\sqrt{\gamma}\,d^2\sigma - \frac{\lambda}{2} \int B_{\mu \nu} \, 
dz^{\mu} \wedge dz^{\nu} \, ,
\eeq
with $\gamma \equiv -{\rm det}\gamma_{a b}$ ($\gamma_{a b}\equiv 
g_{\mu \nu}\,\pa_a z^\mu\,\pa_b z^\nu$), and where the action 
for the fields is 
\beq
\label{5}
S_f = \frac{1}{16 \pi G}\, \int d^n x \sqrt{g}\,\left [
{\cal R}(g) -2\,\co^\mu \vphi\,\co_\mu \vphi 
-\frac{1}{12}\,e^{-4\alpha \vphi}\,H_{\mu \nu \rho}\, 
H^{\mu \nu \rho}\,\right ] \,,
\eeq
with $H_{\mu \nu \rho} = \partial_{\mu} \, B_{\nu \rho} + \partial_{\nu} \, 
B_{\rho \mu} + \partial_{\rho} \, B_{\mu \nu}$, $g \equiv -\det \, (g_{\mu 
\nu})$ (we use the ``mostly plus'' signature). 
Note that $g_{\mu \nu}$ is the ``Einstein'' metric 
(with a $\vphi$-decoupled kinetic term $\sqrt{g}\,{\cal R}(g)$), while 
the ``string'' metric (or $\sigma$-model metric) to which the string is directly 
coupled is $g_{\mu \nu}^s \equiv e^{2\alpha \vphi}\,g_{\mu \nu}$. 
The dimensionless quantity $\alpha$ parametrizes the strength of the coupling 
of the dilaton $\vphi$ to string matter, while the quantity $\lambda$ 
(with same dimension as the string tension $\mu$) parametrizes 
the coupling of $B_{\mu \nu}$ to the string. 
The values of these parameters for fundamental (super)-strings are, in $n$ 
dimensional spacetime, (see, e.g., \cite{DGHR90})
\beq
\label{6}
\alpha_{\rm fs} = \sqrt{2/(n-2)}\,,\quad \quad \lambda_{\rm fs} = \mu\,.
\eeq
Unless otherwise specified we shall, for definiteness, work in $n=4$ dimensions, 
so that $\alpha_{\rm fs}=1$. The additional coupling $\propto e^{-4\alpha \vphi}$
in Eq.~(\ref{5}) between $\vphi$ and the kinetic term of the $B$-field is 
uniquely fixed by the requirement that $\vphi$ be a ``dilaton'' 
in the sense that a shift $\vphi \rightarrow \vphi +c$ be classically reabsorbable 
in a rescaling of the (length and mass) units, i.e. of $g_{\mu \nu}$ 
and the (Einstein-frame) gravitational constant $G$. 

In the present string case the spacetime sources $J(x)$ of the previous 
generic formalism are worldsheet distributed
\beq 
J^i(x)= \left [\frac{\delta S_{\rm int}}{\delta A_i(x)}\right ]_{A=0} = 
\int d^2\sigma\,\sqrt{\gamma^0(z(\sigma))}\,
\delta^{(n)}(x-z(\sigma))\,\bar{J}^i(z)\,,
\eeq
with $\gamma^0=-\mbox{det} 
\gamma^0_{a b}$ and $\gamma^0_{ab} \equiv 
\eta_{\mu \nu}\,\pa_a z^\mu \,\pa_bz^\nu$. 
Inserting this representation into Eq.~(\ref{23}) leads 
($z_1^\mu \equiv z^\mu(\sigma_1)\,, z_2^\mu \equiv z^\mu(\sigma_2)
\,, \gamma^0_1\equiv \gamma^0(z_1)$)
to 
\bea
\label{24a}
&& S_1[z] =\frac{1}{2}\int \int d^2\sigma_1\,d^2\sigma_2 \,
\sqrt{\gamma_1^0}\,\sqrt{\gamma_2^0}\,(4\pi\,G_{\rm sym}(z_1,z_2))\,C_A(z_1,z_2)\,,\\
\label{24b}
&& C_A(z_1,z_2)=\frac{1}{4\pi}\,\bar{J}^i(z_1) R_{i j} \bar{J}^j(z_2)\,.
\eea
The very general formula (\ref{24a}) will be our main tool for clarifying
 the paradoxes raised above. First, in 4 dimensional spacetime, the integral
(\ref{24a}) diverges logarithmically as $\sigma_2^a \rightarrow \sigma_1^a$. 
There are several ways to regularize this divergence. A simple, formal procedure, 
used in the previous literature \cite{CHH90}, \cite{BS}, is to use 
the explicit expression of the 4-dimensional symmetric Green function 
$G_{\rm sym}(z_1,z_2) = 1/(4 \pi)\,\delta((z_1-z_2)^2)$ to perform
the $\sigma_2^0$ integration in Eq.~(\ref{24a}), and then to regularize 
the $\sigma_2^1$ integration by excluding the segment 
$-\delta_c < \sigma_2^1 -\sigma_1^1 <\delta_c$. 
Here, the conformal-coordinate-dependent 
quantity $\delta_c$ is linked to the invariant cutoff $\delta 
\equiv ({\gamma^0})^{1/4}\,\delta_c= \sqrt{\gamma_{11}^0}\,\delta_c$. 
Other procedures are to use the regularized Green function 
$G_{\rm sym}^{\rm reg}(z_1,z_2) = 1/(4 \pi)\,\delta((z_1-z_2)^2+\delta^2)$
\cite{D75},\cite{LundRegge}, or dimensional continuation 
\cite{BD98}. We have checked that these different procedures 
lead to the same results. 
By comparing (\ref{24a}) to the zeroth-order string action $S_0[z]=-\mu(\delta)\,
\int d^2\sigma_1\,\sqrt{\gamma_1^0} $, it is easily seen that the 
coincidence-limit-divergent contribution from (\ref{24a}) generates the term 
$ +\log(1/\delta)\,\int d^2\sigma_1\,\sqrt{\gamma_1^0}\,C_A(z_1,z_1)$ 
which renormalizes $S_0[z]$ when $C_A(z,z)$ is independent of $z$, 
as it will be. In this case, we 
have the very simple link that the $A$-contribution to the renormalization coefficient 
$C$ of Eq.~(\ref{3}) is simply equal to the coincidence limit of 
Eq.~(\ref{24b}): 
\beq
\label{cnew}
C_A = C_A(z,z)= \frac{1}{4\pi} \bar{J}^i(z) R_{i j} \bar{J}^j(z)\,.
\eeq 
This result allows one to compute in a few lines the various $C_A$'s. 
The worldsheet-densities $\bar{J}_\vphi(z)$, $\bar{J}^{\mu \nu}_g(z)$, 
$\bar{J}^{\mu \nu}_B(z)$, of the sources for $\vphi$, $g_{\mu \nu}$ and $B_{\mu \nu}$ 
(linearized around the trivial background (0,$\eta_{\mu \nu}$,0))
are easily obtained by varying Eq.~(\ref{4}) (e.g. 
$J_\vphi(x) = \left [\delta S_s/\delta \vphi(x)\right ]_{\vphi=0} = 
\int d^2 \sigma\,\sqrt{\gamma^0}\,\bar{J}_\vphi(z)\,\delta(x-z)$). 
They read:
\beam
\bea
\label{newa}
&&\bar{J}_\vphi(z) = -2\alpha\,\mu = 
-\alpha\,\mu\,\gamma_\lambda^\lambda \,, \\
&&\bar{J}_g^{\mu \nu}(z) = -\frac{1}{2}\mu\,\gamma^{\mu \nu}\,,\\
\label{newc}
&&\bar{J}_B^{\mu \nu}(z) = - \frac{1}{2}\lambda\,\epsilon^{\mu \nu}\,,
\eea
\eeam
where 
\beq
\label{gamma}
\gamma^{\mu \nu} \equiv \gamma_0^{a b}\,\pa_a z^\mu\,\pa_b z^\nu\,, 
\quad \quad \quad 
\epsilon^{\mu \nu} \equiv \epsilon^{a b}\,\pa_a z^\mu\,\pa_b z^\nu\,,
\eeq 
($\epsilon^{1 0} = -\epsilon^{0 1}= 1/\sqrt{\gamma^0}$)
are the worldsheet metric and the Levi-Civita tensor, viewed from the external 
(background) spacetime. The residue-matrices $R_{i j}$ are also simply obtained 
by writing the (linearized) field equations $\delta S_{\rm tot}/\delta A=0$
in the form $\Box A = -R J$. This yields 
\beam
\bea
\label{rja}
&& R^\vphi\,\bar{J}_\vphi= 
4 \pi G\,\bar{J}_\vphi\,, \\
\label{rjb}
&& R^g_{\mu \nu \rho \sigma}\,\bar{J}^{\rho \sigma}_{g}= 
32 \pi G\,(\bar{J}^g_{\mu \nu} 
-\frac{1}{n-2}\,\eta_{\mu \nu}\,\bar{J}_\lambda^{g\,\lambda})\,, \\
\label{rjc}
&& R^B_{\mu \nu \rho \sigma}\,\bar{J}^{\rho \sigma}_B= 32 \pi G\,\bar{J}^B_{\mu \nu}.
\eea
\eeam
Applying Eq.~(\ref{cnew}) yields, 
in any dimension $n$~\footnote{In $n>4$ dimensions the leading 
ultraviolet divergences are $\propto C\,\delta^{4-n}$ which poses the problem 
of studying also the subleading ones.}  our main results
\beam
\bea
\label{25a}
&& C_\vphi= (G\,\alpha^2\,\mu^2)\,(-2)^2=
+4G\,\alpha^2\,\mu^2\,, \\
\label{25b}
&& C_g = 2G\,\mu^2\,\left [ \gamma_{\mu \nu}
{\gamma}^{\mu \nu} - \frac{(\gamma_\lambda^\lambda)^2}{n-2}\,\right ]
= 4G\,\mu^2\,\frac{n-4}{n-2}\,,\\
\label{25c}
&& C_B = 2G\,\lambda^2\,\epsilon_{\mu \nu}
\epsilon^{\mu \nu}=  -4G\,\lambda^2 \,.  
\eea
\eeam
In the four dimensional case this yields Eqs.~(\ref{17a})--(\ref{17c}). 
Note that $C_g$ vanishes only in 4 dimensions. Note also 
that the sum $C_{\rm tot}=C_\vphi + C_g+C_B$ vanishes for fundamental 
strings (non renormalization \cite{DH89}, \cite{DGHR90}), 
Eq.~(\ref{6}), in any dimension, but that for $n\neq 4$ it is crucial to include 
the non-vanishing gravitational contribution. The special nature 
of the coincidence-limit cancellations taking place for 
fundamental strings is clarified by using, instead of conformal coordinates
$(\sigma^0, \sigma^1)$, null worldsheet coordinates $\sigma^\pm = \sigma^0 
\pm \sigma^1$. Indeed, in terms of such coordinates one finds the simple 
left-right factorized form (typical of closed-string amplitudes) 
\beq
\label{32}
\sqrt{\gamma_1^0}\,\sqrt{\gamma_2^0}\,C_{\rm tot}^{\rm fs}(z_1,z_2) =
32G\,\mu^2\,(\pa_{+}z_1^\mu)\,(\pa_{+}z_{2 \mu})\,(\pa_{-}z_1^\nu)\,(\pa_{-}z_{2 \nu})\,,
\eeq
where $\pa_\pm z^\mu \equiv \pa z^\mu/\pa \sigma^\pm$.
In the coincidence limit, $z_1=z_2=z$, the right-hand side of Eq.~(\ref{32})  
vanishes because $\pa_\pm z^\mu$ are null vectors (the Virasoro constraints 
reading $(\pa_\pm z^\mu)^2=0$).

Using our general result (\ref{24a}) we can now exhibit the link 
between the {\it ultraviolet} object $C=C(z,z)$ and {\it infrared}, 
i.e. long-range, effects. Indeed, let us consider a system 
made of {\it two straight and parallel} 
(infinite) strings (with the same orientation of the axionic source 
$\epsilon^{\mu \nu}$), which are, at some initial time, at rest with respect to each other. 
The condition for this initial state  of relative rest to persist 
is that the interaction 
energy between the two parallel strings be zero, or at least 
independent of their distance. But the interaction energy is just 
(modulo a factor $-2$ and the omission of a time integration) the effective 
action (\ref{24a}) in which $z_1$ runs on the first string, while 
$z_2$ runs on the second one. As, in the case of two straight 
and parallel strings, $C(z_1,z_2)$ is independent of $z_1$ and $z_2$, 
we see that the vanishing of the tension-renormalization coefficient
$C=C(z,z)$ (initially defined as an ultraviolet object)
is equivalent, through the general formula (\ref{24a}), 
to the absence of long-range forces between two parallel strings 
(which is an infrared phenomenon).  
This result allows us not only to make the link with the infrared-based 
arguments of Refs. \cite{DH89}, \cite{DGHR90} and notably 
with the no-long-range force condition discussed in Ref.~\cite{DGHR90} 
(where they find, in 4-dimensions, a compensation between attractive scalar 
forces and repulsive axial ones), but also to understand in simple terms why 
the gravitational contribution to $C$ vanishes: this is simply related to the fact 
that, in 4 dimensions, straight strings exert no gravitational 
forces on external masses. 

Summarizing in symbols, we have shown that 
$C^{\rm effective-action}_{\rm ultraviolet} = 
C_{\rm infrared}^{\rm long-range-force}$. 
We have also independently verified, 
by a direct calculation of the string equations of motion, 
that there were errors in the dynamical calculations of 
Ref.~\cite{CHH90} and that the correct result 
was indeed given by Eqs.~(\ref{25a})--(\ref{25c})
\cite{BD98}, so that, in symbols, 
$C_{\rm ultraviolet}^{\rm dynamical}=
C^{\rm effective-action}_{\rm ultraviolet}$.
As is discussed in detail in Ref.~\cite{BD98}, the main problem with 
the dynamical calculations of Ref.~\cite{CHH90} (besides some computational 
errors for the dilaton force) is that the equations of motion for 
self-interacting strings, without external forces, are sufficient 
to prove renormalizability, but do not contain enough information
for extracting the value of the tension renormalization. 
To determine unambiguously the renormalization 
of $\mu$ one needs, either to explicitly couple the string to external 
(say,  axionic) fields, or to work only with the strictly 
variational equations of motion $\delta S_s/\delta z^\mu$. 

There remains, however, 
to understand the discrepancy between the dynamical $C$'s and the 
expected field-energy ones, 
Eqs.~(\ref{18a})--(\ref{18c}). 
This puzzle is resolved by noting that the coupling of a string to 
$B_{\mu \nu}$ (as well as the coupling of a point particle to $A_{\mu}$ 
considered by Dirac) is the only one to be {\it metric-independent}, 
$S_B^{\rm int} = -\frac{1}{2}{\lambda}\,\int B_{\mu \nu} \,
dz^\mu \wedge dz^\nu$, and therefore the only one not to contribute to the total 
stress-energy tensor $T^{\mu \nu}_{\rm tot} = 
2 g^{-1/2}\,\delta S/\delta g_{\mu \nu}$. By contrast, for the fields 
$\vphi$ and $g_{\mu \nu}$ the total interaction energy cannot be unambiguously 
localized only in the field, there are also interaction-energy contributions which 
are localized on the sources. These (divergent) source-localized 
interaction-energies are included in the effective action 
$S_1[z]$ but are missed in $T^{\mu \nu}_{\rm field-energy}$, thereby explaining 
the discrepancies for $C_\vphi^{\rm field-energy}$ and $C_g^{\rm field-energy}$.

To conclude, let us summarize the new results of this work. We have
derived the ``one-classical-loop'' (i.e. one classical self-interaction)
effective action for Nambu-Goto strings interacting via dilatonic, 
gravitational and axionic fields. Its explicit form, obtained 
by inserting Eqs.~(\ref{newa})--(\ref{newc}) and Eqs.~(\ref{rja})--(\ref{rjc}) into 
Eq.~(\ref{24a}) and Eq.~(\ref{24b}), reads in any spacetime dimension $n$, 
\beq
{S}_c^{\rm eff}[z] = -\mu(\delta)\,\int d^2 \sigma_1\,
\sqrt{\gamma_1^0} + 
\frac{1}{2}\,\int \int d^2 \sigma_1\, d^2 \sigma_2\,
\sqrt{\gamma_1^0}\,\sqrt{\gamma_2^0}\,(4 \pi\, G_{\rm sym}(z_1,z_2))\,
C_{\rm tot}(z_1,z_2)\,, \\
\eeq
where $G_{\rm sym}(z_1,z_2)$ is the symmetric scalar Green function  and 
\beq
C_{\rm tot}(z_1,z_2) = C_{\vphi} + C_g(z_1,z_2) + 
C_B(z_1,z_2)
\eeq
with 
\beam
\bea
&& C_\vphi = 4G\,\alpha^2\,\mu^2 \,, \\
&& C_g(z_1,z_2) = 2G\,\mu^2\,\left [ \gamma_{\mu \nu}(z_1)\,
\gamma^{\mu \nu}(z_2) - \frac{1}{n-2}\, \gamma_{\mu}^{\mu}(z_1)\, 
\gamma^{\nu}_{\nu}(z_2)\right ]\,,\\
&& C_B(z_1,z_2) = 2 G\,\lambda^2\,\epsilon_{\mu \nu}(z_1)\,
\epsilon^{\mu \nu}(z_2)\,.
\eea
\eeam
Here $\gamma^{\mu \nu}(z)$ and $\epsilon^{\mu \nu}(z)$ are the worldsheet 
metric and the Levi-Civita tensor, viewed from the external (Minkowski) 
spacetime, Eq.~(\ref{gamma}). In the special case of fundamental 
strings, Eq.~(\ref{6}), the integrand of the first order contribution 
to the effective action simplifies to the left-right factorized form 
(\ref{32}), when written in terms of null worldsheet coordinates. 
In $4$ dimensions, the coincidence limit ($z_1 \rightarrow z_2$) 
generates logarithmic divergences in the first-order contribution to 
${S}^{\rm eff}_c$ which can be absorbed in a  renormalization 
of the bare string tension $\mu(\delta)$. The explicit 
value of this renormalization is given by Eq.~(\ref{3}) and 
Eqs.~(\ref{17a})--(\ref{17c}). A simple understanding of the physical 
meaning of the various field-contributions to the renormalization 
of $\mu$ has been reached: (i) the values and signs of the various 
contributions are directly related to the worldsheet sources and 
the propagators of the various fields, Eq.~(\ref{24b});
(ii) the effective action approach allows one to relate the long-range interaction energy, 
and thereby the long-range force, between two straight and parallel 
strings to the coefficient $C$ of the logarithmic divergence 
in the string tension. [In particular, this explains in simple 
terms why the gravitational contribution to $C$ vanishes (in 4 
dimensions)]; (iii) the previously emphasized vanishing of the tension 
renormalization coefficient $C$ in the case of fundamental strings
\cite{DH89}, \cite{DGHR90} is clarified in two ways: 
(a) by relating it (following (ii)) to the absence of long-range 
force between parallel fundamental strings [a fact interpretable
in terms of supersymmetric (BPS) states], and (b) by exhibiting 
the new, explicit, left-right factorized form (\ref{32}), which 
clearly vanishes in the coincidence limit because of the Virasoro constraints
[ a fact valid for the bosonic string, independently of any supersymmetry
argument]; (iv) finally, a puzzling discrepancy between 
the signs of the renormalization coefficients expected from 
Dirac's field-energy approach to renormalization, 
Eq.~(\ref{2}) and Eqs.~(\ref{18a})--({\ref{18c}), and the (correct) 
signs obtained by the effective action approach has been clarified by 
emphasizing the necessary existence of source-localized interaction energies
for fields which are not $p$-forms. 

\section*{Acknowledgements}
We thank Brandon Carter and Gabriele Veneziano for instructive
exchanges of ideas.

\end{document}